\documentclass[times,twocolumn,final]{elsarticle}
\pdfoutput=1

\usepackage{framed,multirow,multicol}

\usepackage{amssymb}
\usepackage{latexsym}
\usepackage{ textcomp }

\usepackage{xcolor}

\usepackage{hyperref}

\usepackage{amsmath,amsfonts}

\begin{document}


\begin{frontmatter}

\title{Localized Motion Artifact Reduction on Brain MRI Using Deep Learning with Effective Data Augmentation Techniques}

\author{Yijun Zhao$^a$
}

\author{Jacek Ossowski$^b$
}

\author{Xuming Wang$^a$}

\author{Shangjin Li$^a$}

\author{Orrin Devinsky$^c$}
\author{Samantha P. Martin$^c$}
\author{Heath R. Pardoe$^c$}

\address[1]{Fordham University, 113 60th Street, New York, NY 10023, USA}
\address[2]{QS Investors, LLC, New York, NY 10022, USA}
\address[3]{Grossman Medical School, 145 East 32nd Street, New York, NY 100162, USA}


\begin{abstract}

In-scanner motion degrades the quality of
magnetic resonance imaging (MRI) thereby reducing its utility in the detection of clinically relevant abnormalities.
We introduce a deep learning-based MRI artifact reduction
 model (DMAR) to localize and correct head motion artifacts
in brain  MRI scans. Our approach integrates the latest
advances in object detection and noise reduction in Computer Vision. Specifically, 
DMAR employs a two-stage approach: in the
first, degraded regions are detected using the Single Shot
Multibox Detector (SSD), and in the second, the
artifacts within the found regions are reduced using a convolutional
autoencoder (CAE). We further introduce a set of novel data
augmentation techniques to address the high
dimensionality of MRI images and the scarcity of available data. As a result, our model was trained on a
large synthetic dataset of 225,000 images generated from 
375 whole brain T1-weighted MRI scans. DMAR visibly reduces image artifacts when applied to
both synthetic test images and 55 real-world motion-affected slices from 18 subjects from the multi-center Autism Brain Imaging Data Exchange (ABIDE) study. Quantitatively, depending on the level of degradation, our model achieves a 27.8\%--48.1\% 
reduction in RMSE and a 2.88--5.79 dB gain in PSNR on a
5000-sample set of synthetic images. For real-world artifact-affected scans from ABIDE, our model reduced the variance of image voxel intensity within artifact-affected brain regions (p = 0.014).


\end{abstract}


\end{frontmatter}


\section{Introduction}
\label{sec:introduction}

MRI acquisition often requires extended amounts of time within which patients are asked to remain motionless. 
In-scanner head motion is particularly problematic in children, the elderly and in individuals with neurological disorders (\cite{pardoe2016motion}).    
We propose a deep learning-based MRI artifact
reduction model (DMAR) for retrospective correction of
motion artifacts in brain  MRI scans. Our model targets the
typical “ringing” artifacts caused by in-scanner head motion
during the MRI acquisition. Because these rings appear in
various sections of the images, we design our DMAR model
in two stages. In the first stage, we employ the Single
Shot Multibox Detector (SSD, \cite{liu2016ssd}) to localize the regions with ringing
artifacts. In the second, we train a convolutional autoencoder
(CAE,  \cite{hinton2006reducing}) to reduce the artifacts in the regions
identified by the localizer.
There are two primary challenges associated with our approach:
1) motion-related artifacts cause errors in the initial
time-domain signals, which manifest as spatially extended
distortions, and 2) the high dimensionality of
imaging datasets versus the relatively limited amount of training
data. We address the first challenge by modeling the ringing artifacts as controlled perturbations in the $k$-space representation of an MRI
scan (\cite{likes1981moving, twieg1983k, ljunggren1983imaging}) as well as by modulating the original scan's voxel intensities. The second challenge is addressed by augmentation of a limited number of motion-free MRI scans; we obtain a large number of synthetic motion-free images by applying smooth transformations that
alter proportions of the subjects' morphological features. As a result, we have generated a set of 225,000 artificial images facilitating the training of our deep-learning model.
 We present the details of the image
generation process in Section \ref{sec:augmentation}.

We evaluate our model's performance using both a synthetic dataset of 5000 artificially corrupted images with various degradation levels and a set of scans with visually identified artifacts from the multi-center Autism Brain Imaging Data Exchange (ABIDE) study (\cite{di2014autism}). We use three quantitative measures: the pixel-wise root mean squared error (RMSE) and peak signal to noise ratio (PSNR, \cite{psnrbook}) applied to our synthetic image dataset, and regional standard deviations of image intensities assessed in real-world images where the ground truth was unavailable. 
Details are provided in Section \ref{sec:results}.

The main contribution of our study is the two-stage approach to reduce motion artifacts in MRI brain scans. Existing methodologies typically apply a correction model to the entire image. Our experiments indicate that this nonselective approach can overcompensate in regions where the artifacts are less prominent. Moreover, the two models we developed in each stage are independent; they can be integrated (as in our DMAR model) or applied to separate tasks where each could be of value. Our work also introduces a set of new augmentation techniques to generate large numbers of realistic MRI images (both motion-free and motion-affected). 
DMAR was effectively trained on synthetic data, demonstrating the feasibility of our approach and its potential to improve image quality in clinical imaging environments.

\section{Related Work}
\label{sec:related}
In the area of artifact localization, Lorch et al. studied detection of motion-affected regions using a supervised learning approach based on random decision forests \cite{lorch2017automated}. Both the effects of respiratory motion on cardiac scans and of bulk patient motion in the acquisition of head scans were studied. 
Kustner et al. provided a method for spatially resolved detection of motion artifacts in MR images of the head and abdomen \cite{kustner2018automated}. In their study, images were divided into partially overlapping patches of different sizes achieving spatial separation. Using these patches as input data, a convolutional neural network (CNN) was trained to derive probability maps for the presence of motion artifacts in the patches. The authors concluded that identifying motion artifacts in MRI is feasible with good accuracy in the head and abdomen. 

To the best of our knowledge, we are the first to apply a deep object detection model (i.e., SSD) to localize motion artifacts in brain MRI scans. Our experimental results demonstrate the great practical utility of the approach as evidenced by the model's high mAP score (Section \ref{sec:mAP}).

On the subject of MR image denoising, 
the majority of current studies can be classified into two categories:  image-based and $k$-space-based approaches. Typical image-based techniques include super resolution (\cite{gholipour2010robust, peled2001superresolution, chaudhari2018super}), convolutional neural networks (\cite{zhang2017beyond, hauptmann2019real, schlemper2017deep, pawar2018motion}), and generative adversarial network (GAN) models (\cite{ lyu2020mri,    yang2017dagan, jiang2019respiratory}). For example, Zhang et al. introduced a deep convolutional network which reduced image artifacts resulting from Gaussian noise with unknown noise level \cite{zhang2017beyond}. The authors showed a way to apply their model to other reconstruction problems such as single image super resolution and JPEG image deblocking. 
Hauptmann et al. applied a temporal convolutional neural network to reconstruct highly accelerated radial real-time data of cardiovascular magnetic resonance images \cite{hauptmann2019real}.
Lyu et al. proposed an approach to improving the quality of MR images using the ensemble of five generative adversarial networks (GANs), each of which working with a dataset produced by a different conventional image super resolution method \cite{lyu2020mri}. They found that the ensemble outperformed any single sub-network and produced results superior to those of other deep learning-based super-resolution methods. 

$k$-space-based models constitute a parallel vein of popular methods in medical image processing (\cite{pawar2019deep, oksuz2019automatic, shaw2020k, akccakaya2019scan, hyun2018deep, terpstra2020deep, han2019k, lee2019k, kim2019loraki}). 
For example, Pawar et al. applied deep learning methods to the problem of reconstruction of artifact-free MRI scans from their randomly undersampled $k$-spaces \cite{pawar2019deep}. They transformed the problem to that of pixel classification, and solved it using a deep learning classification network. 
Oksuz et al. proposed a method to classify low quality cardiac magnetic resonance (CMR) cine images by detecting the presence of motion-related artifacts in such images \cite{oksuz2019automatic}. To address the limited availability of low-quality samples, the authors generated artificial motion-corrupted images by replacing $k$-space lines in a high-quality image frame with lines from other temporal frames. Shaw et al. introduced another $k$-space-based technique for generation of synthetic motion-corrupted samples to facilitate training of deep learning models \cite{shaw2020k}. In their approach, the artifact-free input volume was resampled according to a randomly sampled movement model, defined by a sequence of ``demeaned" 3D affine transforms. These 3D Fourier transforms were subsequently combined to form a composite $k$-space, which was transformed back to the image domain producing the final artifact volume.

Our study belongs to the second category. We introduce a new method of generating realistic motion artifacts in brain MR images using controlled perturbation of the $k$-space of motion-free frames. Specifically, we modify the $k$-space data of a motion-free image by selecting a symmetric, centered annular sector and applying a uniform phase shift and scaling of the annular sector's elements (Section \ref{sec:kspace}). We further supplement our $k$-space-based approach by applying elliptic intensity modulations on motion-free slices to generate the ``rippling" artifacts (Section \ref{sec:rippling}). In contrast to existing studies, we also generate synthetic artifact-free images by modeling natural inter-subject variability in brain morphology (Section \ref{sec:warp}). These data augmentation techniques allow us to significantly increase the number of instances in training our deep learning models.

\section{Material and methods}
\label{sec:material}

We utilized a two-stage process for correcting motion-related ringing artifacts in structural MRI scans (Figure 1). All analyses were carried out on individual MRI slices. Image artifacts were first localized using the Single Shot Multibox Detector (SSD), and the identified regions  were then corrected using a denoising convolutional autoencoder (CAE). The CAE model was trained by following the supervised learning paradigm in which a set of corrupted images is provided and the network learns to reconstruct the corresponding artifact-free images which are provided as the ground truth. Since both models require large amount of training data, augmentation techniques were used to generate a large set of artifact-free slices and their corresponding artifact-corrupted counterparts.

\begin{figure}[t]
\vspace{2mm} 
\includegraphics[ trim = 0 200 0 50, clip, scale=0.32]{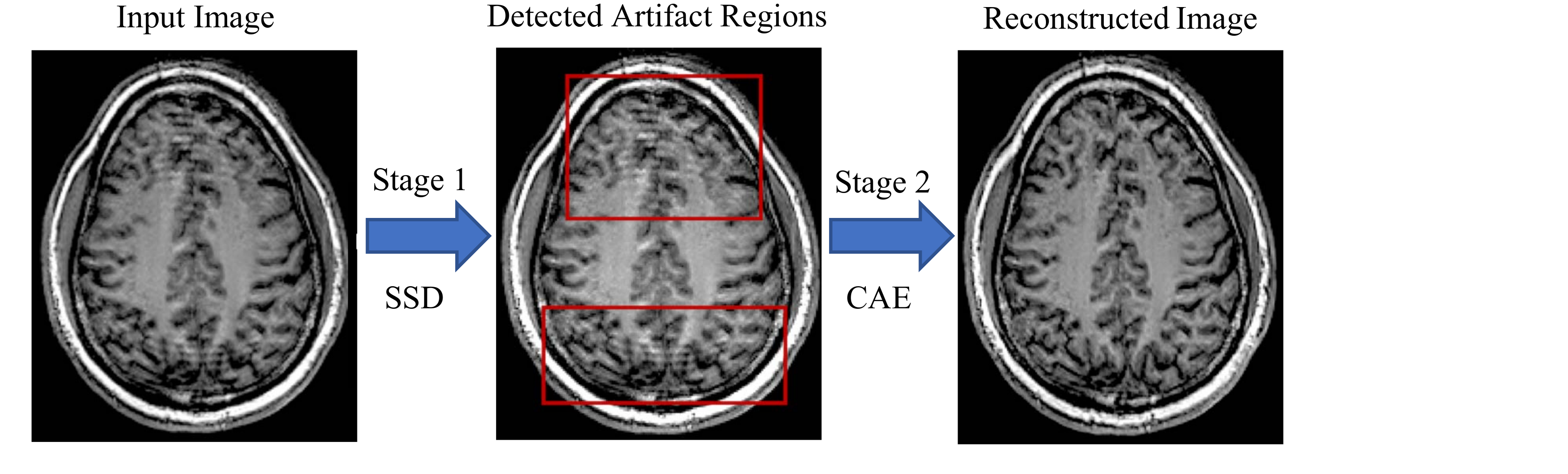}
\caption{Two-stage Approach to Reduction of Motion Artifacts. SSD: Single Shot MultiBox Detector. CAE: Convolutional Autoencoder.
}
\label{fig:intro}
\end{figure}

\subsection{ Data Acquisition}
\label{sec:data}

\subsubsection{Model Training}
\label{sec:data:train}

Our  large synthetic  image dataset was generated using the OASIS-1 \cite{marcus2007open} dataset. OASIS-1 contains 436 T1-weighted MRI scans of 416 subjects (Age: 52.7\textpm 25.1; F/M: 61.5\%); 20 subjects had two MRI sessions. All scans were selected through a per-slice screening process along each principal axis to ensure their quality. Of these, 375 scans from 355 subjects were used to generate our training data, and the remaining 61 scans from 61 subjects were held out for model testing.

To generate the training data, 50 slices were randomly sampled along each of the sagittal, axial, and coronal directions from each MRI scan. Thus, a total of $150 \times 375=56,525$ slices were used to generate the synthetic motion-free images as well as their corrupted versions with localized image artifacts (Section \ref{sec:augmentation} ). A total of 225,000 pairs were generated to facilitate training of our artifact localization and reduction models.

\subsubsection{Model Testing}
\label{sec:data:test}

The model was tested on both synthetic and real-world MRI slices. A synthetic test set with 5000 image pairs was generated from the held out set of 61 subjects following the same process as in the model training. To study our model's performance at various levels of corruption, 1000 images were randomly selected in each of the following degradation intervals as indicated by PSNR: $<$17, $[17, 18)$, $[18, 19)$, $[19, 20)$, and $[20, 21)$. The set was used for quantitative evaluation of our model's performance using PSNR and RMSE measures.

\begin{figure*}[t]
\vspace{2mm} 
\centering \includegraphics[ trim = 100 530 70 80, clip, scale=1]{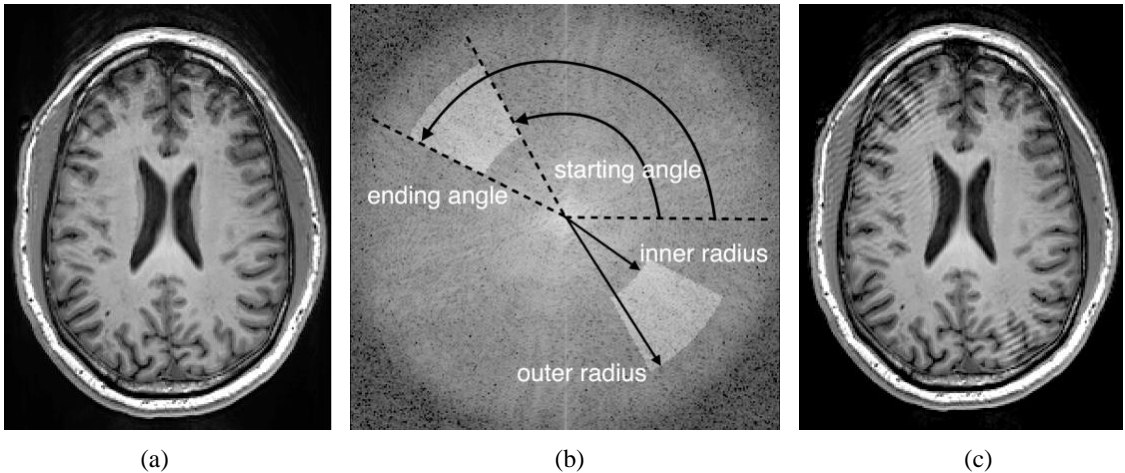} 
\caption{ $k$-space Based Artifact Generation. (a) Original image; (b) Symmetric, centered annular sector is modified in the $k$-space of the image; the sectors' elements are uniformly scaled and their phase is shifted by the same angle; inverse FT is applied to obtain an artifact image (not shown) (c) Synthetic image with ``ringing" artifacts obtained by superimposing the original image  with a random collection of circular regions from the artifact image constructed in (b).}
\label{fig:kspace}
\end{figure*}

The model was further applied to 18 T1-weighted MRI scans selected from the ABIDE study (\cite{di2014autism}). These scans were selected from a larger dataset that had been visually evaluated as low quality in a previous study (\cite{pardoe2016motion}). Image slices from the three cardinal planes (axial, coronal and sagittal) were used to validate our model as measured by the standard deviations of image intensity within motion-affected regions identified using the SSD (Section \ref{sec:cae-eval2}). Qualitative visual assessment was performed on both test datasets.

\subsection {Data Augmentation}
\label{sec:augmentation}

Three augmentation methods were used to generate our training data for both stages of the DMAR model. To learn detecting the artifacts, the localizer required a set $C$ of artifact-free images, a set $D$ of the degraded versions of images in $C$, and a set $B$ of collections of bounding boxes locating the artifacts in $D$. The autoencoder model used the same sets $C$ and $D$ to learn the reconstruction function which produced the best approximation of a motion-free image from a corrupted one. To accommodate these needs, our methods produce i) a large dataset of realistic motion-free images based on a limited number of real-world scans (Section \ref{sec:warp}), and ii) a corresponding set of corrupted images with localized ringing artifacts (Sections \ref{sec:kspace} and \ref{sec:rippling}). These synthetic artifacts were modeled on real-world motion-affected MRI scans from the ABIDE database. The ABIDE images were visually inspected to determine the types of degradations that are likely to occur in real-world imaging datasets. Based on these images, we simulated the ``ringing" and ``rippling" artifacts which we combined in the 2:1 proportion in our final dataset.

\subsubsection {Modeling inter-subject brain morphological variability}
\label{sec:warp}

We applied local spatial distortions to simulate natural inter-subject variability in brain morphology. These deformations were performed in a varying set of three to eight non-overlapping circles within each motion-free image. The number and location of the circles changed randomly from image to image and their radii were chosen to be maximal while still allowing no overlaps (resulting in frequent tangent pairs). Within each circle, a radial stretching was applied with a smoothly changing ratio that equaled 1.0 both at the circle's center and its border. This ensured that the created deformations were localized, had no discontinuities, and blended smoothly with the unaffected areas. The stretching ratio varied according to the formula:\\

                                                   IMG$_{new} (P)$ = IMG$_{old} ( C + u^{(1+\epsilon)} (P - C) )$\\
                                                   
\noindent where   IMG$_{new} (P)$ is the new pixel intensity at a given point $P$ in a circle with the center $C$ and radius $R$, and $u = distance(P, C)/R$. We have found that setting $\epsilon$=0.2 resulted in a moderate amount of deformation and provided great variability between the images. The obtained transformations are quite subtle in that an untrained observer may not notice them unless the images are viewed in quick succession (e.g., in an  
\href{https://storm.cis.fordham.edu/~yzhao/100_distortions_BW.mp4}{\color{blue}animation}). This process allowed us to generate hundreds of different images from a single MRI slice and ultimately to create a large data set of images.

\subsubsection{ k-space based synthetic artifact generation}
\label{sec:kspace}
Raw MRI data is encoded in $k$-space, representing the spatial frequencies of the object being imaged. The $k$-space data is then converted into the human-recognizable MRI scan by the application of an inverse Fourier transform (FT). In-scanner head motion during the scan introduces errors as $k$-space is filled that manifest as ringing, ghosting or blurring artifacts following the inverse FT. Therefore artifacts similar to those encountered in clinical imaging can be generated by modifying the FT of a motion-free image in $k$-space and applying an inverse FT. We modified the $k$-space data of a given image by selecting a symmetric, centered annular sector and applied a uniform phase shift and scaling of the magnitude of the annular sector's elements (Figure \ref{fig:kspace}). 

We utilized three  sets of base parameters (Table 1) to create synthetic artifacts that resembled real-world artifacts in images from the ABIDE study.  In each of the base sets three subgroups of the parameters, a) inner and outer radii, b) modulus magnification, c) starting and ending angles of the annular sectors, were randomly shifted by multiples of 2-3 pixels, 0.2, and 11$^{\circ}$-29$^{\circ}$ respectively. This resulted in the annular areas' uniformly covering the whole mid-to-high frequency section of the Fourier spectrum as they varied from image to image.

\renewcommand{\arraystretch}{1.2}
\begin{table}[!h]
\caption{ Base Combination of Annular Parameters}
\label{tab:parameters}
\centering
\small
\begin{tabular}{ | r | c | c | c | c |}
\hline
\bf{Parameter} & \bf{Comb. 1} & \bf{Comb. 2} & \bf{Comb. 3}  \\
\hline
inner radius (pixels) & 61  & 55 & 60 \\
\hline
outer radius (pixels) & 66 & 105 &80  \\
\hline
phase shift & 144$^\circ$  &29$^\circ$ & 9$^\circ$\\
\hline
modulus magnification & 9  & 8 & 6 \\
\hline
annular sector starting angle &  10$^\circ$ & 29$^\circ$ & 11$^\circ$ \\
\hline
annular sector ending angle & 38$^\circ$ & 57$^\circ$ & 46$^\circ$ \\
\hline
\end{tabular}
\end{table}

Following these $k$-space operations, we applied an inverse FT to convert to image space and obtain an artifact image that suffered degradation across its whole area. To localize the corruption,  we copied a random collection of circular ROIs with random sizes and locations from the artifact image onto the original clear scan to obtain a set of localized artifacts. The final distorted image was intensity histogram matched with the original image to prevent pixel intensity shifts. We thereby generated in this fashion hundreds of motion-affected images from a single high quality slice. In addition, our random number generator settings ensured that the centers, radii, angles, frequencies, and magnitudes we employed in sizing and positioning of the artifacts were different on every synthetic image.

\begin{figure*}[!t]
\vspace{2mm} 
\includegraphics[ trim = 30 280 0 0, clip, scale=0.6]{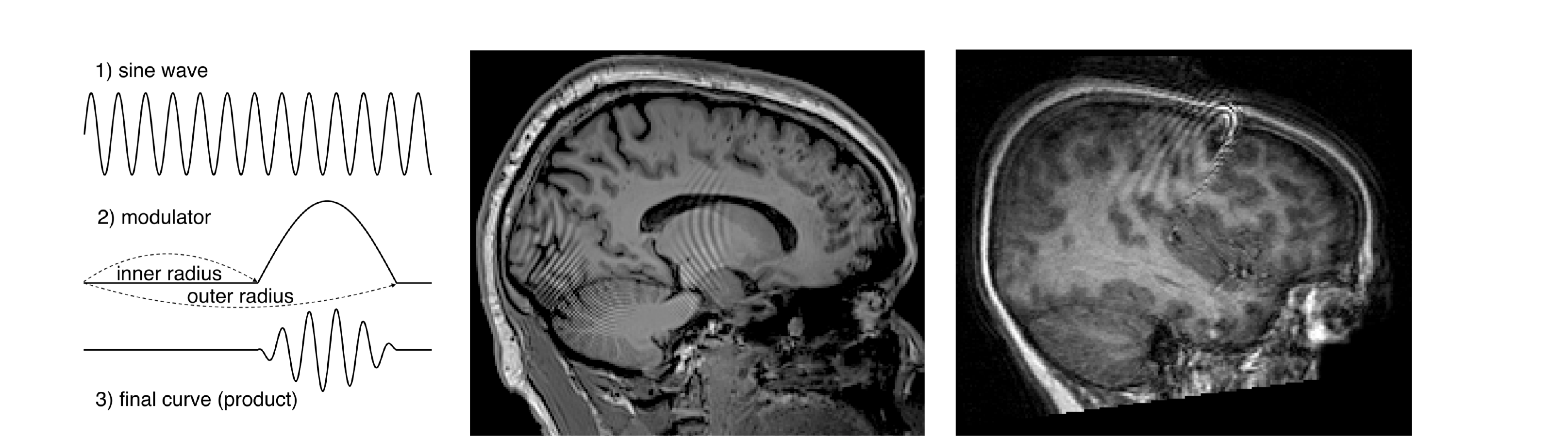}
\caption{ Synthetic Elliptic Artifact Generation. (a) Three-step process for generating elliptic intensity variations. $S$ is a sine wave that propagates elliptically from a randomly chosen center in a given image. $M$ is a modulator curve consisting of the positive part of the sine function extended between random inner and outer radii. Pixel intensities in the image are multiplied by $1 + M*S$ ; (b) sample synthetic artifact; (c) sample real-world elliptic artifact.}

\label{fig:rippling}
\end{figure*}

\subsubsection {'Rippling' artifact generation}
\label{sec:rippling}
Some MRI scans contain artifacts that appear as spatially localized elliptic artifacts that resemble a rippling effect (Figure \ref{fig:rippling}). We supplemented the approach described in \ref{sec:kspace} by applying elliptic intensity modulations on motion-free slices. The effect was obtained by following a three-step process (Figure \ref{fig:rippling} (a)): 1) generating a sine wave $S$ that propagated elliptically from a randomly chosen center point, 2) generating a modulator curve $M$ which consisted of the positive part of the sine function extended between random inner and outer radii, and 3) multiplying the scan's pixel intensities by $1 + M*S$. The frequency of the initial wave and the amplitude of the modulator were chosen randomly giving rise to hundreds of artifacts per original image. As described in Section \ref{sec:kspace}, to further localize the degradation, a random set of circular ROIs was extracted from the artifact image and superimposed over the clear scan (Figure  \ref{fig:rippling}(b)). We present an example of real-world scan manifesting such artifacts in Figure \ref{fig:rippling}(c).

\subsection{ Deep MRI Artifact Reduction (DMAR) Model }

\begin{figure*}[!t]
\centering  \includegraphics[ trim = 15 20 0 0, clip, scale=0.55]{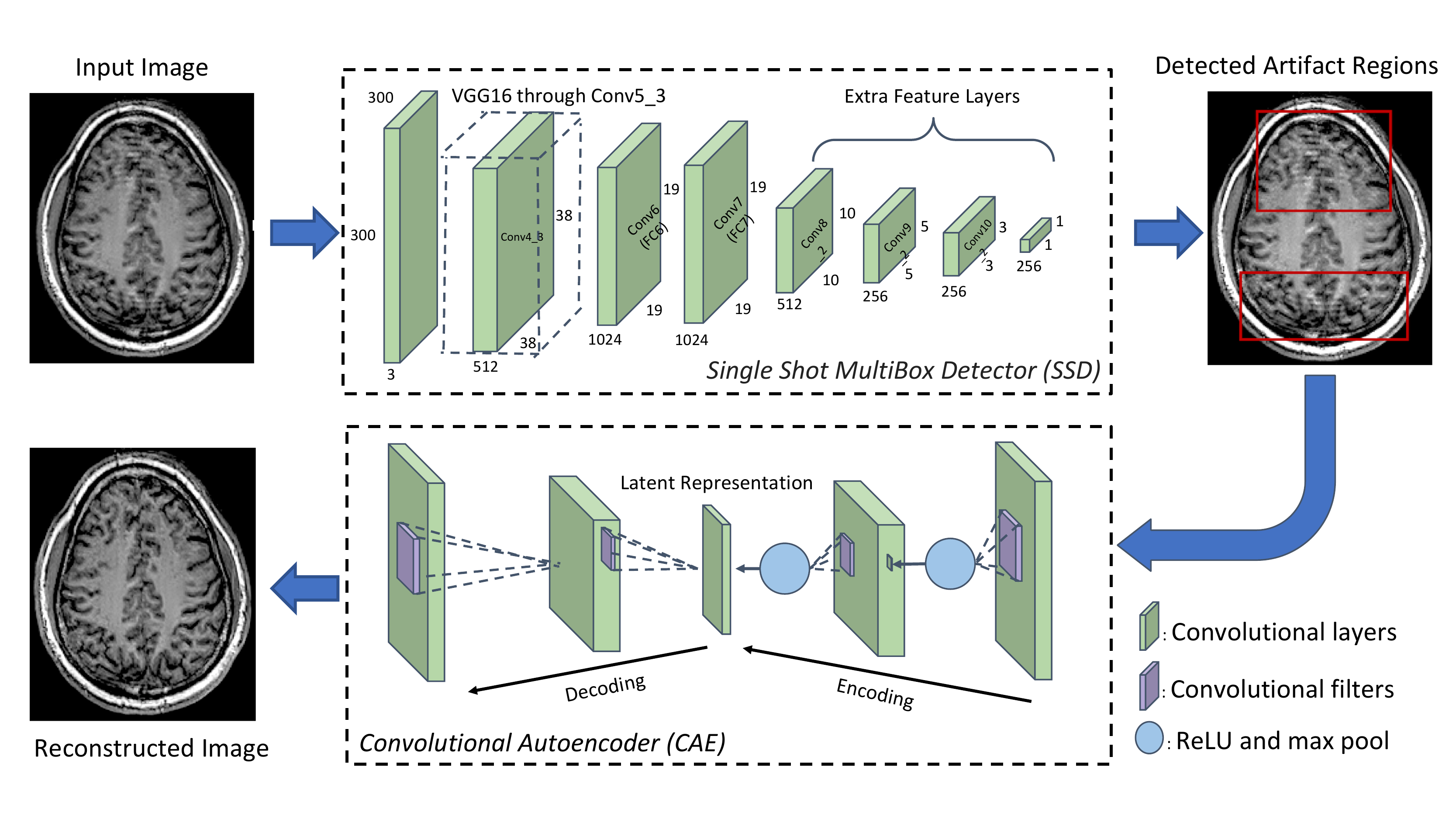}
\caption {DMAR Model Architecture. SSD outputs detected regions with artifacts as indicated by the red boxes. CAE reconstructs the image within the boxed regions.}
\label{fig:dmar}
\end{figure*}

Figure \ref{fig:dmar} presents the two-step pipeline of our DMAR model. 
To implement the localization component (upper diagram), we investigated different state-of-the-art models including the Single Shot MultiBox Detector (SSD, \cite{liu2016ssd}) and Mask R-CNN (\cite{he2017mask}), and selected SSD for its better accuracy and computational efficiency. We briefly describe the SSD model and our customizations in Section \ref{sec:ssd}. In the second step (bottom part of the diagram), DMAR applies a convolutional autoencoder (CAE) to reduce the artifacts within the regions identified by the localizer. The CAE network and our customizations are introduced in Section \ref{sec:autoencoder}.

\subsubsection{Single Shot MultiBox Detector (SSD) }
\label{sec:ssd}
Unlike other object detection systems (\cite{he2017mask, ren2015faster, girshick2015fast}) where the object localization (i.e., bounding box positions) and classification are carried out by two separate networks, SSD (\cite{liu2016ssd}) accomplishes both tasks in a single network, which is computationally more efficient and easier to integrate into systems that require a detection sub-system. The SSD internal architecture consists of a base convolutional neural network (CNN) for extracting relevant image features and additional feature layers to allow for object detection at differing spatial scales.  
We experimented with two established CNN networks that are often applied in their entirety or as components to computer vision problems: the VGG (\cite{simonyan2014very}) and ResNet (\cite{he2016deep}), ultimately settling on VGG-16.

The SSD outputs (i) bounding boxes indicating the location of detected artifacts, and (ii) a score between 0 and 1 indicating the likelihood that each bounding box corresponds to an object of the specified type. For our study, we trained the SSD to detect a single object type, i.e., an artifact, with multiple occurrences possible per slice. We trained our SSD model using the synthetic datasets described in Section \ref{sec:data:train} with a 4:1 split for training and validation. The input to the model were the images superimposed with localized artifacts. The ground-truth bounding boxes were the circumscribing squares of all ROIs generated during the data augmentation process. 
The evaluation of our SSD implementation is presented in Section \ref{sec:ssd-eval}.

\subsubsection{ Convolutional Autoencoder (CAE) }
\label{sec:autoencoder}

A CAE is a deep learning approach that combines the power of a convolutional neural network (CNN, \cite{zeiler2014visualizing}) and an autoencoder (\cite{hinton2006reducing}). The former extracts defining features from images and the latter has been widely applied to data compression and image noise reduction. Integration of the two approaches has delivered promising results in tasks such as object recognition (\cite{he2016deep, russakovsky2015imagenet}), image captioning (\cite{karpathy2015deep, pu2016variational}) and image restoration (\cite{zhang2017learning, mao2016image}).

Our model is inspired by a variant of autoencoder, the {\it denoising} autoencoder (\cite{xie2012image, vincent2008extracting}), which takes as input a set of degraded images and is forced to output the corresponding artifact-free images. In the training, both the MRI images corrupted with motion artifacts as well as  their clear versions are provided following the standard supervised learning scheme.

\begin{table}[!t] 
\centering
   \caption{Architecture of Our CAE Model }
   \label{tab:arch}
   \vspace*{2mm}
   \small
  \begin{tabular}{|r|c|c|c|}
    \hline
             & Filter Size  & Stride & \# of Filters\\   
    \hline
    Convolutional Layer1 & 4x4& (1,1)&64 \\
     \hline

    Convolutional Layer2 & 16x16& (1,1) & 128\\
     \hline

    Convolutional Layer3 & 1x1& (1,1) & 256\\
     \hline

    Transpose Conv. Layer 1 & 8x8& (1,1) & 32\\
     \hline

    Transpose Conv. Layer 2 & 1x1& (1,1) & 1\\

  \hline
\end{tabular}

\end{table}

There are five hidden layers in our network. The size, stride, and number of filters for each layer are presented in Table \ref{tab:arch}. 
We train the model using the synthetic dataset (Section \ref{sec:data:train}) with a 4:1 split for training and validation.
The evaluation of the CAE component is presented in Sections \ref{sec:cae-eval1} and \ref{sec:cae-eval2}.

\subsection {Quantitative Efficacy Measures}
\label{sec:metric}

\subsubsection{Localization Model}
\label{sec:mAP}

\begin{figure}[!t]
\includegraphics[trim = 60 180 20 100, clip, scale=0.4]{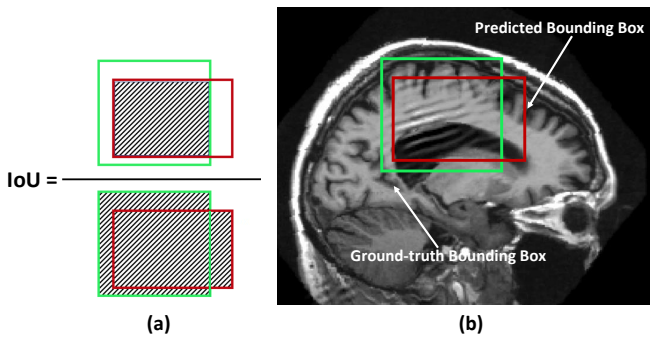}\\
\caption{Intersection Over Union. (a) An IoU is the ratio between the intersection and the union of two areas. (b) A higher IoU ratio indicates a more accurate prediction.}
\label{fig:iou}

{\small }
\end{figure}

To perform well an object detector needs to excel in 1) determining the location of objects (i.e., a regression task), and 2) deciding the type of located objects  (i.e., a classification task). The quality of localization is typically measured by the Intersection over Union ratio (IoU) (Figure \ref{fig:iou}(a)). This ratio captures the alignment of the predicted bounding boxes with those of the ground truth. A high IoU ratio indicates a more accurate prediction. Figure \ref{fig:iou}(b) shows an example of ground-truth (green) and predicted (red) bounding boxes identified by an object detector. We require an IoU value$>$0.5 to endorse a true detection. Varying the IoU threshold alters the detection sensitivity of the model.

Given an IoU threshold, the quality of classification for a single class is measured by the average precision (AP) across a spectrum of recall values (\cite{fawcett2006introduction}). For multiple classes, the AP values are further averaged over all possible classes leading to the mAP measure, which was first formalized in the PASCAL Visual Objects Classes(VOC) challenge (\cite{everingham2010pascal}). mAP scores take values in the interval [0, 1] where 1 indicates a perfect detection.

\subsubsection{DMAR model}
\label{sec:dmar-metric}
Since our DMAR model was trained according to the supervised learning paradigm where the degraded and underlying artifact-free images are available, it is natural to measure its efficacy on separate pairs of clear and corrupted images that did not participate in the training. To this end, we generated from 61 held-out subjects separate image sets $C$ and $D$ (clear and degraded) as described in Section \ref{sec:data:test}. For every degraded image $d \in D$, its corrected version $\text{DMAR}(d)$, and the ground truth image $c \in C$, we measured the similarity between $\text{DMAR}(d)$ and $c$, and that between $d$ and $c$, hoping that the former pair's coupling was tighter than the latter's. Specifically, we applied two similarity measures: pixel-wise root mean squared error (RMSE) and the peak noise to signal ratio (PSNR, \cite{psnrbook}). A smaller RMSE indicates higher similarity between the images. PSNR is defined as the ratio between the maximum possible power of a signal and the power of corrupting noise that affects the fidelity of its representation. A higher PSNR indicates a higher quality of an image. The measures were applied after scaling pixel intensities of the images to the interval [0, 255]. 

Our model was also applied to artifact-affected scans from the ABIDE study. Since the ground-truth images were unavailable, the model was evaluated by comparing the regional variability in image intensity between un-corrected and corrected images within bounding boxes identified in the localizer stage. We hypothesize that the variability in image intensity in regions corrected using our DMAR approach, as measured using the regional standard deviation, would be reduced in our corrected scans. MRI data from 18 subjects scanned at five sites were used in this analysis. We applied the DMAR model to axial, coronal and sagittal slices. Overall 55 slices were analyzed. We compared the within-box standard deviation of corrected and uncorrected scans using a paired T-test.

\begin{figure}[!h]

\centering \includegraphics[ trim = 105 585 140 65, clip, scale=0.67]{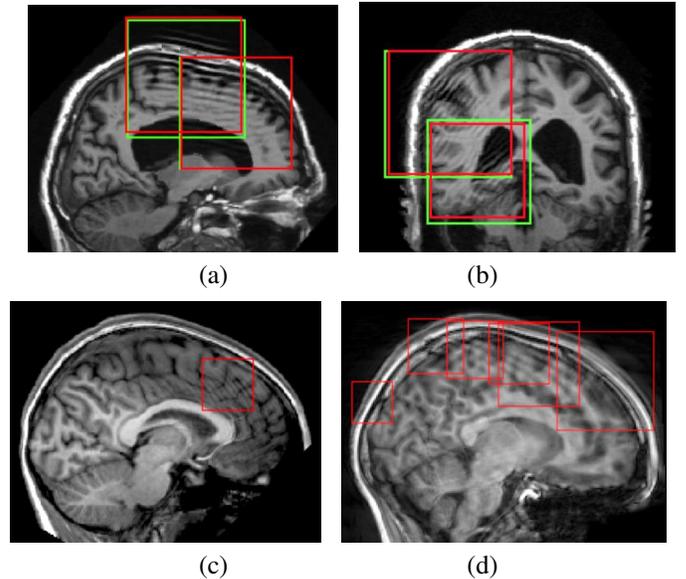}
\hspace{1cm} (a) \hspace{3cm} (b) \\
\vspace{1.5mm}
\centering \includegraphics[ trim = 105 400 0 257, clip, scale=0.67]{localization_final2_oasis.pdf}
\hspace{1cm} (c) \hspace{3cm} (d) 
\caption {Sample Output of the Localization Model on Synthetic Images. Red boxes are the ``ringing" artifact regions detected by the model.
(a) \& (b): synthetic test data with the ground-truth (green boxes) locations.
(c) \& (d): real-world test data without the ground-truth. 
}

\label{fig:localization}
\end{figure}

\section{Results}
\label{sec:results}

In this section we present the experimental results of our model using both synthetic and real-world data with metrics defined in Section \ref{sec:metric}.

\begin{figure*}[!t]
\centering \includegraphics[ trim = 60 375 60 120, clip, scale=1]{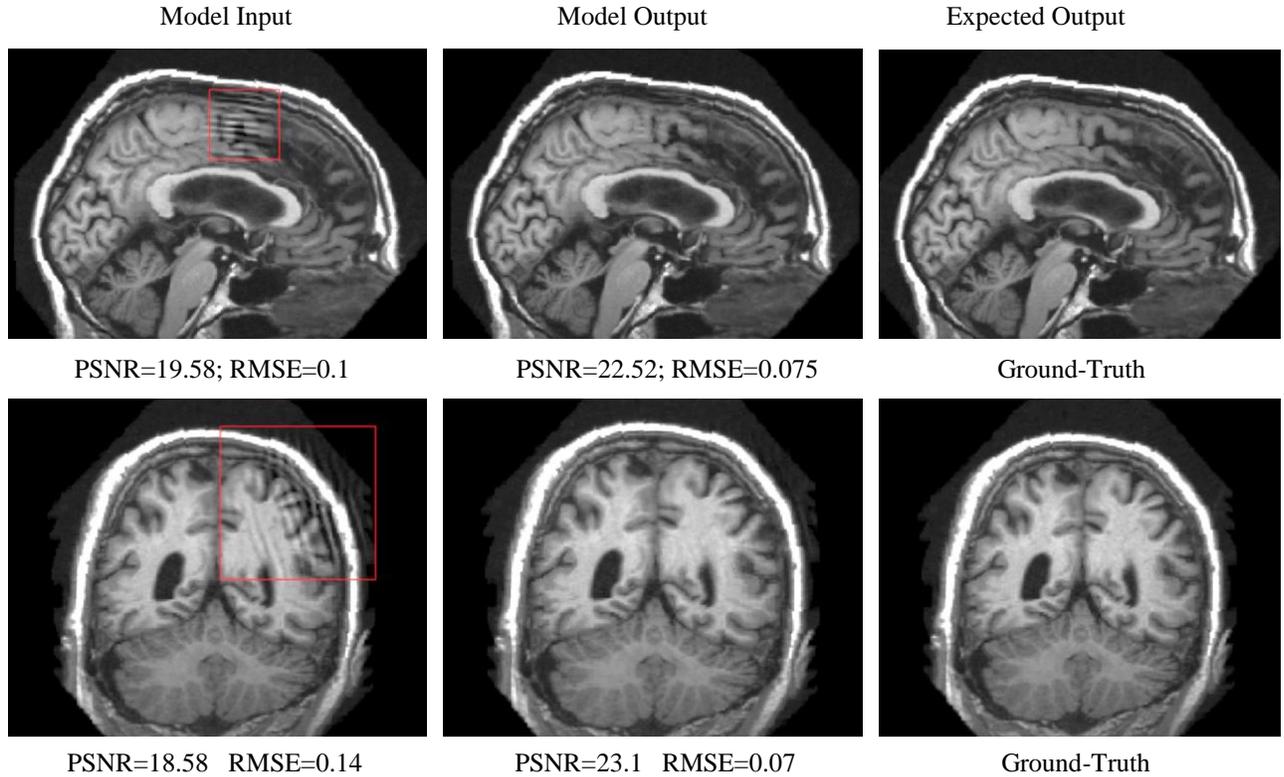}
\caption {Visual Assessment of the DMAR Model on Synthetic Data. 
Rectangular boxes are the artifact regions detected by our localization model. Artifact reductions are performed only within the identified regions. Corrected boxes are visually indistinguishable from the corresponding areas in the ground-truth image.
RMSE and PSNR values are shown for the original and the reconstructed images.}
\label{fig:synthetic}
\end{figure*}

\subsection{ Performance of the Artifact Localization Model }
\label{sec:ssd-eval}

We evaluated DMAR's localization component on the synthetic test set described in Section {\ref{sec:data:test}. Our model achieved an mAP score of 0.88 at IoU level of 0.5.

Qualitatively, our observations indicate that the localization model identifies the ringing artifacts well in practice (Figure \ref{fig:localization}). In the two examples of synthetic images ((a) \& (b)), green boxes indicate the ground-truth locations of the artifacts and red boxes the model's detections. For the two real-life images ((c) \& (d)), ground truth is unavailable, and only predicted locations are shown.

\renewcommand{\arraystretch}{1.2}
\begin{table*}[!h]
\caption{Quantitative Evaluation Across Different Degradation Levels }
\label{tab:quantitative}
\begin{center}
\small
\begin{tabular}{|c| c | c | c | c | c | c|}
\hline
 \multirow{2}{*} { PSNR Level} &\multicolumn{3}{c|} {Pixel-wise RMSE}  & \multicolumn{3}{c|} {PSNR (dB)} \\ \cline{2-7}
    & {Degraded vs. Target}& {Corrected vs. Target}  &  {Reduction(\%)*}  & {Degraded vs. Target}& {Corrected vs. Target}  &  {Gain*}  \\ \cline{1-7}

$<17$ &  0.171 (0.020) &  0.089 (0.016) & {\bf 48.1\%} & 15.35 (0.99) &  21.14 (1.47) & {\bf 5.79}\\

$[17, 18)$&  0.133 (0.004) & 0.078 (0.014) & {\bf 41.4\%} &17.52 (0.29) & 22.31 (1.58) & {\bf  4.79}\\

$[18, 19)$& 0.118  (0.004) & 0.074 (0.012) & {\bf 37.5\%} &  18.50 (0.29) &  22.72  (1.47) & {\bf 4.21}\\

$[19, 20)$ & 0.106 (0.004) &  0.072 (0.010) & {\bf 31.4\%}  & 19.49 (0.29) & 22.85 (1.25) & {\bf 3.36}\\

   $[20, 21)$  & 0.094 (0.003) & 0.068 (0.007) & {\bf 27.8\%} &  20.49 (0.28) & 23.38 (0.99) & {\bf 2.88}
\\ 

\cline{2-4}

\hline

\end{tabular}
\end{center}

The "Degraded vs. Target" columns contain the discrepancies (RMSE) and similarities (PSNR) between corrupted scans and their artifact-free counterparts in each category. The "Corrected vs. Target"  columns contain the discrepancies/similarities between DMAR-corrected images and the targets. The numbers in parentheses represent standard deviations. The values were computed  after first scaling the images to the range [0, 255]. \\

\vspace{-2mm}
*All reductions and gains are statistically significant with t-statistic$>$40.
\end{table*}

\subsection {Evaluation of the DMAR Model on Synthetic Images}
\label{sec:cae-eval1}

We applied the DMAR model to our synthetic test set of 5000 images and compared the output to the ground truth. 
For the analysis, similarities between the following pairs of images were quantified: degraded input vs. target, and model-corrected vs. target. Figure \ref{fig:synthetic} presents DMAR's action on two instances of such pairs. 

Similarity was quantified using RMSE and PSNR, and the improvements produced by the model were measured in \% and dB respectively. A good performance would be indicated by the similarity of the second pair being higher than that of the first. Table \ref{tab:quantitative} presents our model's performance across a spectrum of five degradation levels. Each category contained 1000 images whose PSNR values with respect to the ground truth were within the indicated intervals. This corruption scale was motivated by the model's progressively stronger intervention when faced with increasingly motion-affected images. We observed that for images with relatively small artifacts, PSNR $> 21$dB, DMAR refrained from making substantial corrections, while for those with PSNR $< 19$dB it intervened aggressively producing large improvements. One can clearly see this monotonic trend in the "Gains" column through both RMSE and PSNR metrics.

All reductions and gains in Table \ref{tab:quantitative} are statistically significant with t-statistic$>$40. In our experiments, we continue to observe smaller but still substantial average gains when the input images have PSNR $> 21$dB. 
We have found that images in this category are in general visually close to the ground-truth (e.g., row \#2, center image in Figure \ref{fig:synthetic}) and, thus, not the focus of our study.

\begin{figure*}[!t]
 \hspace{1cm} Model Input \hspace{2.5cm} Model Output \hspace{2.5cm} Model Input \hspace{2.5cm} Model Output

\centering \includegraphics[ trim = 20 55 25 30, clip, scale=0.7]{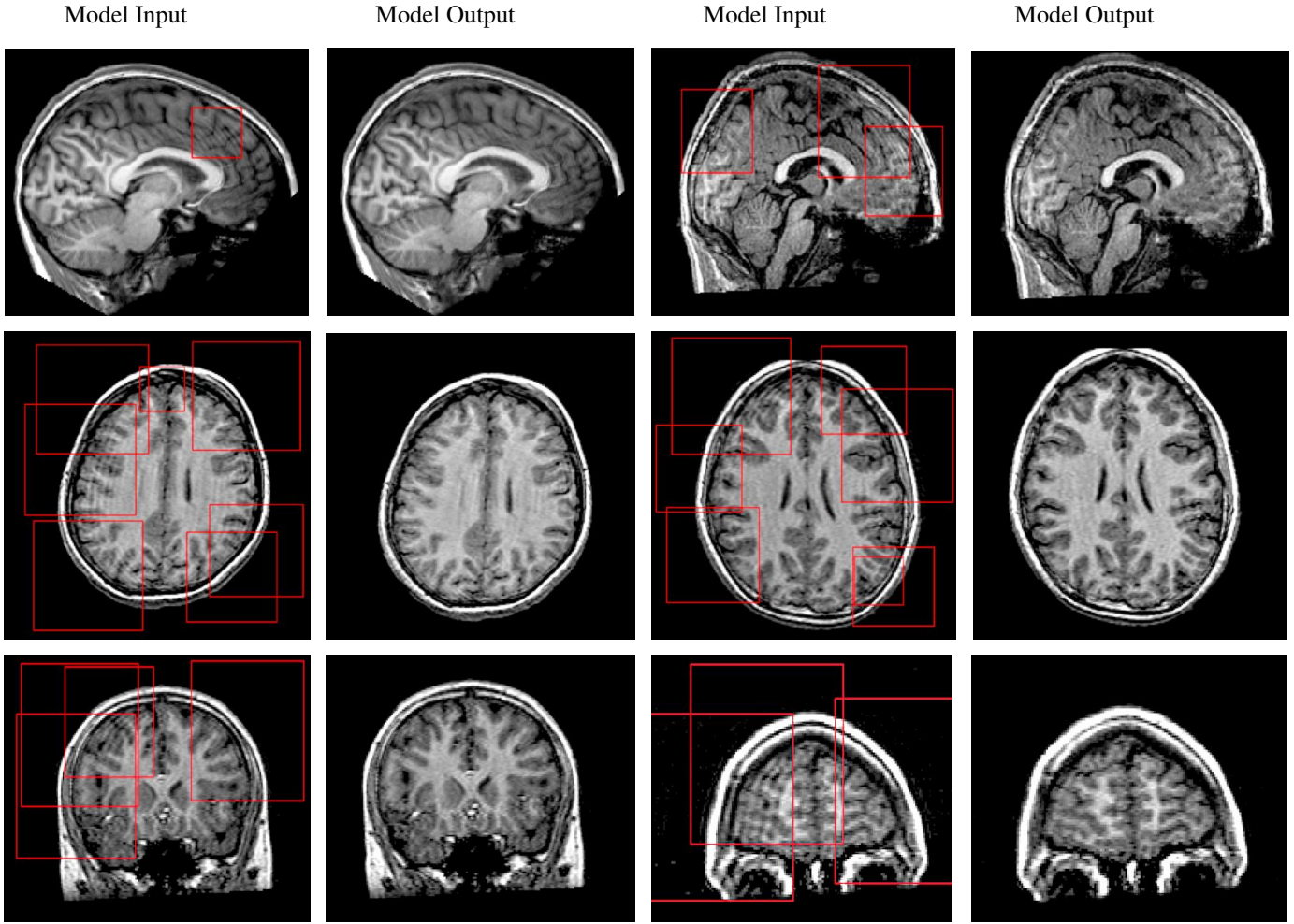}

\caption {Visual Assessment of DMAR Model on Artifact-affected MRI Scans from the ABIDE Study. Red boxes are the detections from our localization model. Motion artifact reductions are performed only within the identified regions. }
\label{fig:real}
\end{figure*}

We view the model's increasing conservatism on images with progressively smaller artifacts as a desirable property which makes DMAR  preserve areas with no corruption within the bounded boxes. This effect was evident on numerous examples of particular images, both synthetic and real-world, where our model did not intervene when prompted with practically clear scans.

\subsection{Evaluation of DMAR model on real world brain MRI scans}
\label{sec:cae-eval2}

We applied our model to a selection of artifact-affected ABIDE MRI scans as described in Section \ref{sec:data:test}. 
Examples of the model output when applied to such scans are shown in Figure \ref{fig:real}. Our quantitative analyses indicate that the image intensity variability  within  bounding boxes identified  by DMAR was reduced in the corrected slices by 0.76\% (p = 0.014 ).

\section{Conclusion }
Our deep learning-based method can reduce ringing artifacts in  brain MRI scans. Our DMAR model integrates the latest advances in Computer Vision, applying deep neural networks to object recognition and image reconstruction. To overcome the scarcity of training data, we introduced techniques in data augmentation and generated large quantities of realistic synthetic brain MRI images. Our methods generate both clear scans as well as images affected by ringing artifacts. The evaluation of DMAR on synthetic datasets showed substantial improvements as measured by PSNR gains and reduction in RMSE. In addition, our model reduced the variability of image pixel intensities in the neighborhoods of ringing artifacts and demonstrated compelling visual improvements in qualitative inspections. These results were obtained using a relatively low number of high quality scans.  They convincingly support the utility of deep learning in reducing image artifacts in brain MRI scans due to in-scanner head motion.

Our approach is limited by the use of two-dimensional slices rather than three dimensional volumes of the patients' scans. In principle, the methods presented here could be extended to the 3D domain by manipulating the 3D versions of the relevant data structures. Although computationally challenging, working with volumes of voxels could result in further improvements in the quality of the reconstructed images.


\bibliographystyle{model2-names}

\bibliography{refs}

\end{document}